\documentclass[aps,preprint,showpacs,showkeys]{revtex4}
\usepackage{amssymb,amsmath,epsf,graphicx}
\usepackage[ansinew]{inputenc}

\begin{document}
\title{Microscopic Details of the Integer Quantum Hall Effect in an Anti-Hall Bar}

\author{Christoph Uiberacker}
\email{christoph.uiberacker@unileoben.ac.at}

\author{Christian Stecher}
%\email{christian.stecher}

\author{Josef Oswald}
%\email{josef.oswald@unileoben.ac.at}

\affiliation{Institute of Physics, University of Leoben, 
Franz-Josef-Strasse 18A, 8700 Leoben, Austria}

\begin{abstract}
Due to the lack of simulation tools that take into account the actual geometry of complicated quantum Hall samples there are lots of experiments that are not yet fully understood.
Already some years ago R.~G.~Mani recorded a shift of the Hall resistance transitions to lower magnetic fields in samples of a Hall bar with embedded anti-Hall bar by using partial gating. We use a Nonequilibrium Network Model (NNM) to simulate this geometry and find qualitative
agreement. Fitting the simulated resistance curves to the experimental results we can not only determine the carrier concentration but also obtain an estimate of the screened gating potential and especially the amplitude and lengthscale of potential fluctuations from charge inhomogenities which are not easily accessible by experiment.
\end{abstract}
\maketitle

\newpage 

\section{Introduction}

The main application of the classical Hall effect lies in the simultaneous determination of the carrier concentration and the mobility. For small enough temperatures and pure samples quantum effects become important. In the quantum Hall regime experimental results for magnetic fields corresponding to plateau transitions depend on microscopic details like the potential landscape, which complicates interpretation.

In experiments of the quantum Hall effect (QHE) a fixed voltage or current is applied to two metallic contacts, while additional contacts are used to measure longitudinal and transversal potential differences. For interpretation it is therefore essential to know the distribution of the electrochemical potential in the sample. Generic properties of this potential for simple geometries have been analysed in a series of papers \cite{MacDonald1983,Chklovskii1992,Guven2003}. The plateaus at multiples of $h/e^2$ of the transversal (Hall) resistance as a function of magnetic field could be explained in terms of a simple modification of the Landauer model by Büttiker \cite{Streda1987,Buttiker1988}, using non-interacting electrons in an empirical confinement potential \cite{Datta1995}. However, the resistance between plateaus depends on the detailed geometry, electrostatics of electrons or the potential landscape generated by excess charges in the doped semiconductors in vicinity to the 2-dimensional electron gas (2dEG).
 
Electrostatics was investigated self-consistently in simple geometries \cite{Chklovskii1992,Siddiki2003} and shown to lead to the formation of alternating compressible and incompressible stripes. One concludes that in the plateau regime current in response to (transversal) electric field only flows in the incompressible stripes.
The picture becomes more complicated for nonideal contacts due to potential barriers or when applying gatings because channels are (partially) blocked and the electrochemical potential is changed in their surroundings. F.~Dahlem et.~al.~found experimentally that width and magnetic field values of the transition region between plateaus can change significantly \cite{Dahlem2010}. At the same time we investigated such situations with the nonequilibrium network model (NNM) \cite{Uiberacker2009,Oswald2006}, with good agreement to the experiments.

In addition the NNM has proven successful for exotic sample geometries like anti-Hall bars within Hall bars \cite{Oswald2005}, when compared to experiments of ungated samples by R.~G.~Mani \cite{Mani1996jpsj,Mani1996epl2}.
Shortly after, Mani applied partial gating to his samples and recorded a shift of the Hall resistance transitions to lower magnetic fields as a function of the gating voltage \cite{Mani1997apl}. In the present paper we describe simulations of these samples with the NNM and show that the features in the experiment are well reproduced.

Moreover, by manually fitting the transversal resistance we extract the electron concentration, the screened gating potential and the average curvature of saddle points of the potential. From the curvature and a statistical model we obtain estimates for the amplitude and lengthscale of potential fluctuations.

\section{Theory}
\label{s_theory} 
 
The exact Hamiltonian of a typical integer quantum Hall sample is complicated due an interface of two semiconducting layers, excess charges, and the presence of metallic contacts. Accurate calculations for the whole structure by ab initio methods fail due to the prohibitively high number of electrons in the heterostructure. 

For sufficiently high magnetic field electron wavefunctions are highly confined to the magnetic length $l_B=\sqrt{\hbar/eB}$, where $e$ and $B$ denote elementary charge and magnetic induction normal to the 2dEG.
Therefore we model the potential landscape in the 2dEG phenomenologically, using a network of semiclassical trajectories in a (slowly varying) random potential that mimics the effects of the excess charges. The most prominent of this type of models is the Chalker-Coddington model \cite{Chalker1988}, which is able to predict statistics of states and scaling exponents but cannot describe the nonequilibrium steady state.
In contrast, our NNM is designed to calculate these nonequilibrium quantities. The model rests on the local equilibrium approximation \cite{Zubarev1974}, which is applied to a network of semiclassical wavefunctions. In this way we attribute unique thermodynamical quantities such as the electrochemical potential to each single wavefunction.

Transport by highly localized quantum states can be viewed as a percolation problem, where it is well known that only pivotal edges are relevant to the (bond) percolation problem \cite{Grimmett1999}. These edges correspond to saddle points of the potential landscape, which then form the network \cite{Chalker1988}.
At each saddle 4 chemical potentials meet and we assume that phases are destroyed by decoherence, such that the distances between saddles become irrelevant.
In this approximation we can replace the potential fluctuation by the model potential $V(x,y)=V[\cos(\omega y)-\cos(\omega x)]$ of a regular grid of saddle points, where the period $L:=2\pi/\omega$ and amplitude $V$ should be understood as averages of the microscopic potential.

We define the ratio of longitudinal to transversal field component at a saddle point as
\begin{equation}
	P := \frac{E_x}{E_y}=\frac{u_1-u_2}{u_1-u_4}=\frac{u_4-u_3}{u_2-u_3} \quad .
\end{equation}
In this way we construct a ''transfer'' equation for the chemical potentials
\begin{equation}
	\left[\begin{array}{c} u_2 \\ u_3 \end{array}\right] =
		 \left[\begin{array}{cc} 1-P & P \\ -P & 1+P \end{array}\right]
		 \left[\begin{array}{c} u_1 \\ u_4 \end{array}\right] \quad .
\end{equation}
The chemical potential distribution can be calculated as a boundary value problem once the values for $P$ at each node are given. We model external contacts supplying current in the NNM by saddles with a pair of trajectories that point into/out of the sample. One of the two states is fixed to the chemical potential of the current supply contact.

We neglect nonlinear effects, that is, the values of $P$ do not depend on the chemical potentials. This should be well justified in case of typical experimental currents of $\mu A$, corresponding to voltage differences lower than $0.1mV$ across the sample or energies well below the Landau level (LL) spacing of $\hbar eB/m^*\approx 1.728meV$ at $B=1T$, respectively.

Except in a small region around the hotspots (at opposite edges of the current inducing contacts) the local conductance tensor can be approximated as purely off-diagonal. This leads immediately to $P=I_y/I_x$. This approximation has the advantage that only the electric field has to be calculated.
According to the edge channel picture we call $T$ the probability of transmission in longitudinal direction, therefore we get $I_y\propto R$ and $I_x\propto T=1-R$.
$P$ can then be calculated from elastic tunneling transition probabilities at the Fermi energy across a saddle as \cite{Fertig1987,Buttiker1990} 
\begin{equation}
	P = \delta_{mn}\frac{R_{mn}}{1-R_{mn}}= \exp\left[-\epsilon\frac{B}{c}\right] \quad ,
	\label{E_def_P}
\end{equation}
with $\epsilon := E_F - E_n - V_S$ the energy of the trajectory, in terms of the difference of the Fermi energy to the energy of state $n$, relative to the saddle energy $V_S$. $V_S$ reflects the sum of all potentials (disorder, confinement...). We define the ''center of the transition'' as $\epsilon=0$. While $B$ denotes the magnetic field strength, the parameter $c:=\frac{hV}{eL^2}$ is related to the curvature of the model potential energy fluctuations with amplitude $V$. 
Similar to other approaches \cite{Guven2003,Nachtwei1999} we calculate longitudinal and transversal resistance from the potential distribution by identifying the current with the macroscopic current direction.

We use LLs and a simple self-consistent Thomas-Fermi approximation where  the potential at zero temperature is obtained from the charge density by multiplication with $C=50mV/(10^{11}cm^{-2})$ (units $e=1$). 
Furthermore we use a constant broadening of $0.5meV$ to mimic scattering by a short range disorder potential. 
Equilibration among edge channels is taken into account by assuming tunneling and using an exponential function with a decay parameter.  
   
\section{Results}

\subsection{Transversal resistance and fit of experiment}

\begin{figure}[htbp]
	\centering
	\includegraphics[width=0.5\linewidth]{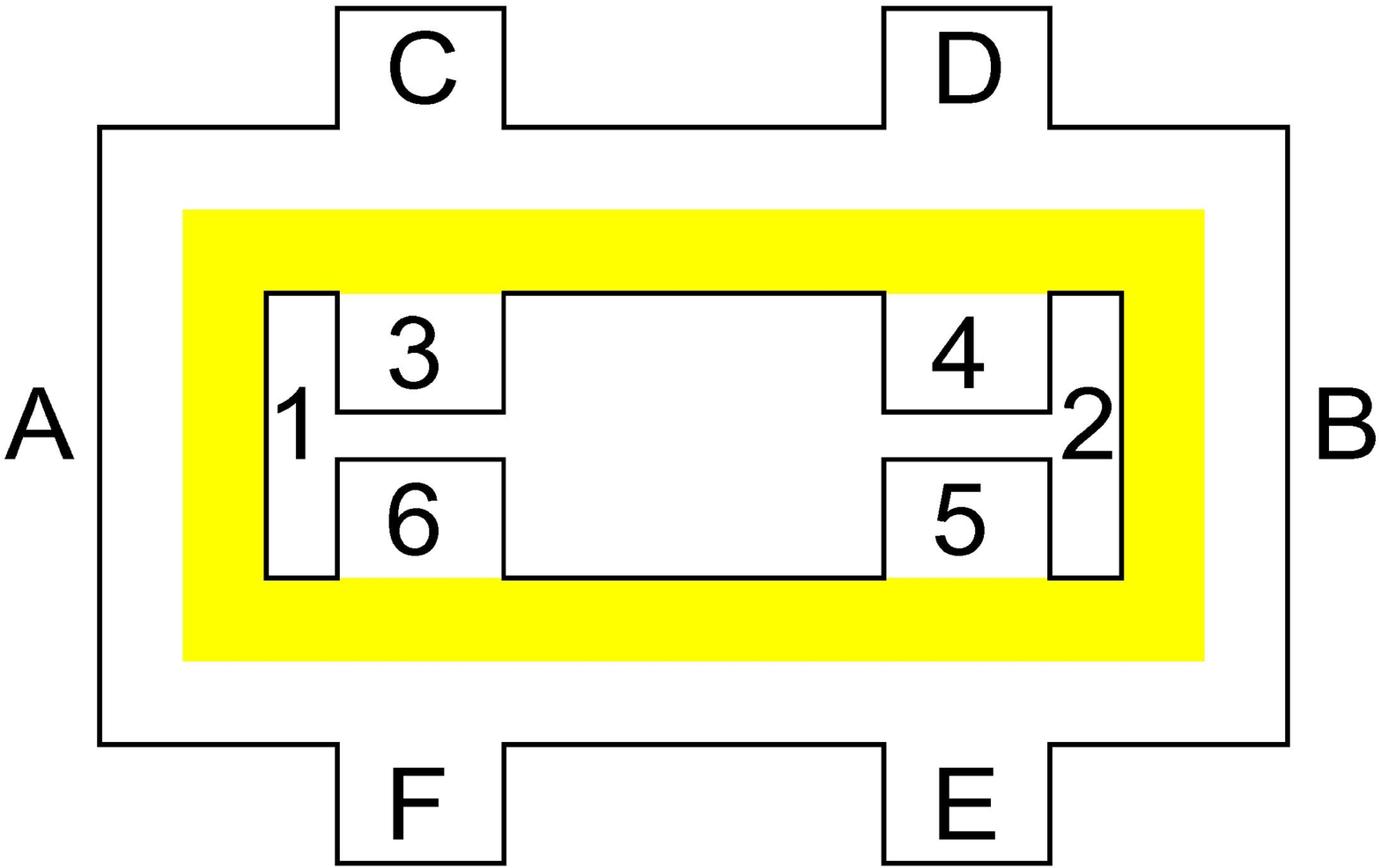}%
%	\caption{The sample as used in the simulation}%
	\caption{\label{F_sampleAntiHall} (color online) Geometry of the anti-Hall bar within a Hall Bar: 
		The figure shows the labelling of contacts used and the shaded, yellow area denotes the partial 
		gating applied as in the experiment.}
\end{figure}

Figure~\ref{F_sampleAntiHall} schematically shows the geometric setup of an anti-Hall bar embedded in a Hall bar, used in the experiment \cite{Mani1997apl}. The gated rectangular region is filled with yellow color and a confinement potential develops near the inner and outer border of the 2DEG, indicated by solid lines. In the experiment a constant inner/outer current was driven through the device by applying appropriate voltage differences at the contact pairs \textit{A-B} of the outer Hall bar and \textit{1-2} of the inner anti-Hall bar. 
We denote in the following the longitudinal (parallel to direction \textit{A-B}) and transversal direction by $x$ and $y$, respectively.
%Longitudinal resistance values $R_{xx}$ are calculated from voltages measured at the pairs of contacts \textit{C-D} , \textit{F-E} as well as
%\textit{3-4}, \textit{6-5} which are then divided by the current measured at {\it A} and {\it 1}, respectively. In the same way 
Transversal Hall resistances $R_{xy}$ are obtained from \textit{F-C} and \textit{6-3}. As discussed in \cite{Uiberacker2009} it is sufficient to use point contacts in simulations of macroscopic Hall samples, such as in the present experiments.

We manually fit our simulations to the experimental results in order to obtain important microscopic information like the carrier concentration, the enhancement factor for the Zeeman energy, the screened gating potential, and the magnitude and correlation length of the potential fluctuations. We estimate errors from half the thickness of lines in plots of the experimental results.

To get the carrier concentration we fit the plateau transition center of $R_{xy}$ of the outer ungated Hall system for the (3 last integer) transitions from filfactor $\nu=4$ to $\nu=3$, $\nu=3$ to $\nu=2$, and $\nu=2$ to $\nu=1$. This results in a carrier concentration of $n = 1.81*10^{11} \pm 0.02*10^{11} cm^{-2}$. Our value is slightly below the range $2*10^{11}cm^{-2}\le n\le 3*10^{11}cm^{-2}$ proposed in the experimental work \cite{Mani1997prb}, where however a fit to the classical Hall slope is normally used.

Due to spin polarization a Zeeman term adds to the energy, resulting in an energy difference of $\Delta E_Z = g\mu B/2$, with $g$ an enhancement factor and $\mu := g_{GaAs}\mu_B$. Here $g_{GaAs} = -0.44$ is the Land\'e factor of GaAs heterostructures and $\mu_B = e\hbar / 2m^*$ the Bohr magneton for electrons with effective mass $m^*$, which we set to the typical effective mass of GaAs, $m^* = 0.067m$. By manually fitting the transition centers in case of no gating we arrive at $g = 12.5 \pm 1.5$. Such large enhancement values are typically seen in transport measurements \cite{Huang2002}.

\begin{figure}[htbp] %
  \centering
		\includegraphics[width=0.5\linewidth]{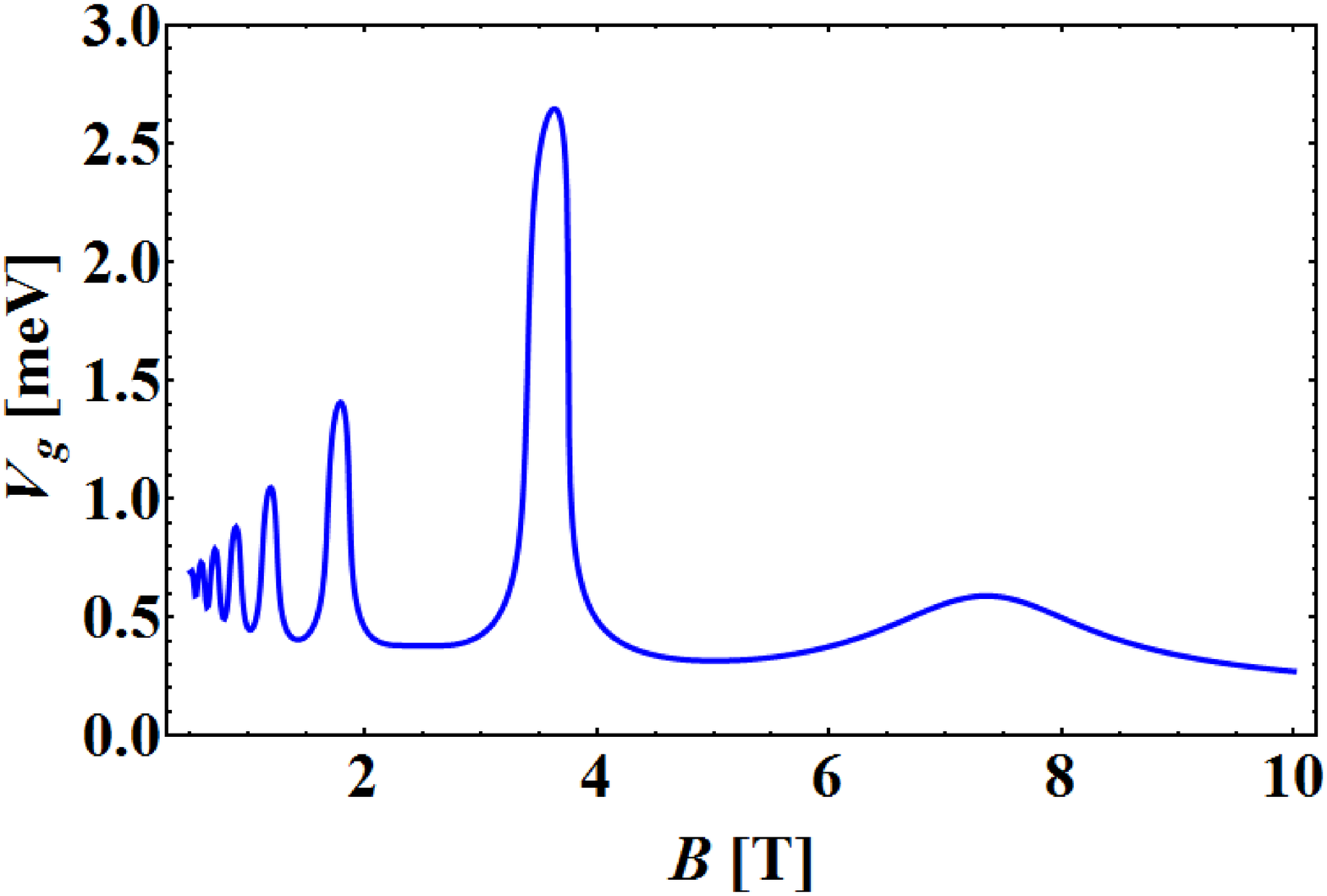} \\
		\includegraphics[width=0.5\linewidth]{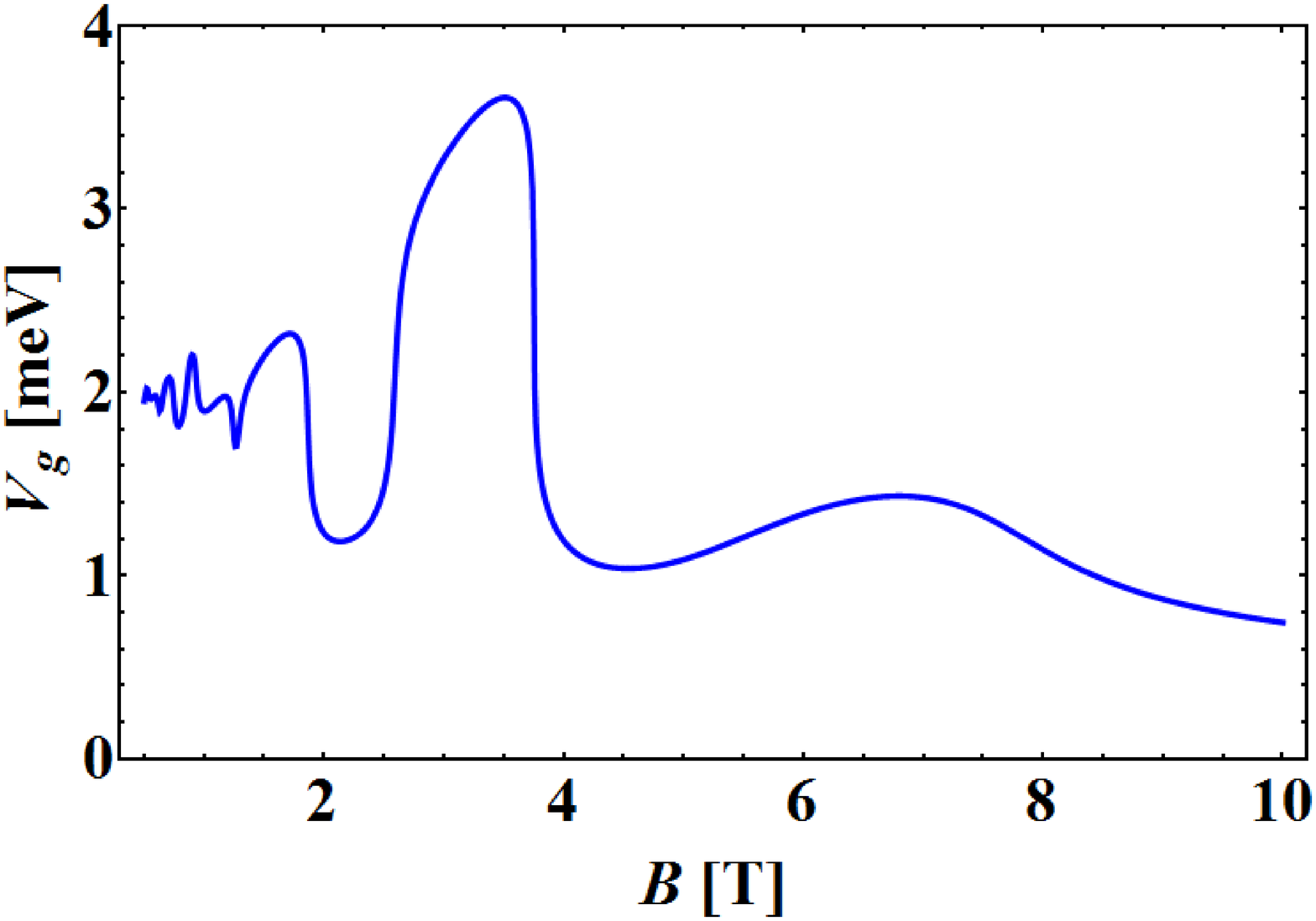} \\ 
		\includegraphics[width=0.5\linewidth]{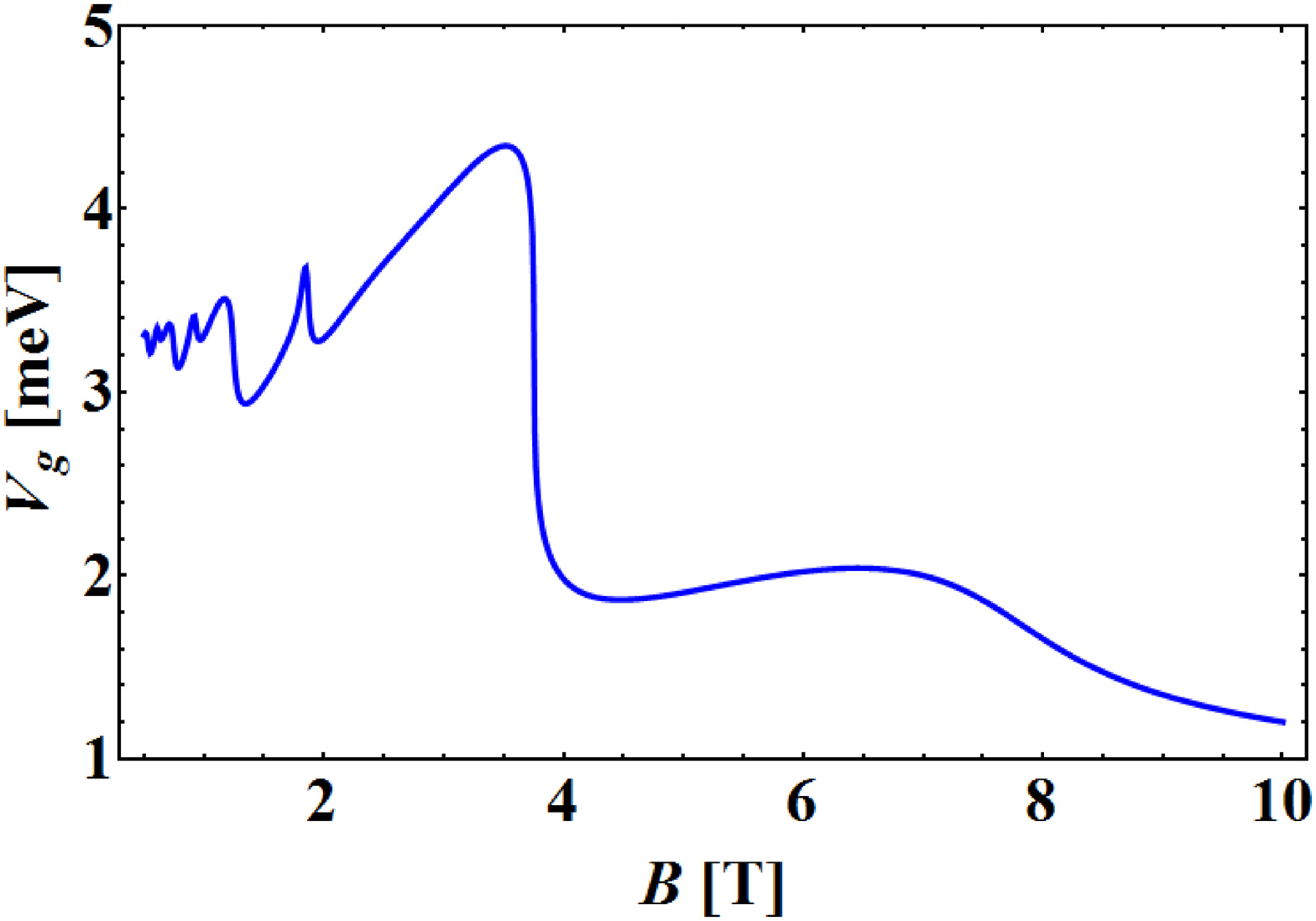}
 \caption{\label{F_Vg}(color online) Screened gating potentials as a function of magnetic field: 
 	We show results for bare gatings of $10meV$, $30meV$, and $50meV$. Low/high screening does not 
 	directly correspond to transition/plateau regions of $R_{xy}$ as broadened LLs for different 
 	spin significantly overlap for the Zeeman energy found.}
\end{figure}

We fit the magnetic field at the center of the plateau transitions observed in the inner and outer leg of the anti-Hall structure, and arrive at bare gating potentials of $10meV$, $30meV$ and $50meV$. 
The screened gating potentials $V_g(B)$ as functions of the magnetic field $B$ are shown in Fig.~\ref{F_Vg} for these bare gating potentials. We note that within the plateau transition $V_g$ is small and fairly constant only for small values of $V_g^0$. 
The magnetic fields at the transition centers are then collected in the table \ref{T_trcenters}.

\begin{table}
	\begin{tabular}{c|c|c|c|c}
		transition 	& $V_g^0 = -450mV$	& $V_g^0 = -300mV$	& $V_g^0 = -150mV$ & $V_g^0 = 0mV$ \\
		\hline
		$2\to 1$ 		&	2.83						& 4.00					& 4.90				& 5.05	\\
		$3\to 2$ 		&	1.55						& 2.19					& 2.72				& 2.93	\\
		$4\to 3$ 		&	1.24						& 1.70 					& 2.20				& 2.26	\\
	\end{tabular}
	\caption{\label{T_trcenters} Magnetic fields in $T$ of the transition centers for various bare gating 
		potentials. We estimate the error as $0.05T$.}
\end{table}

\begin{figure}[htbp]
	\centering
		\includegraphics[width=0.5\linewidth]{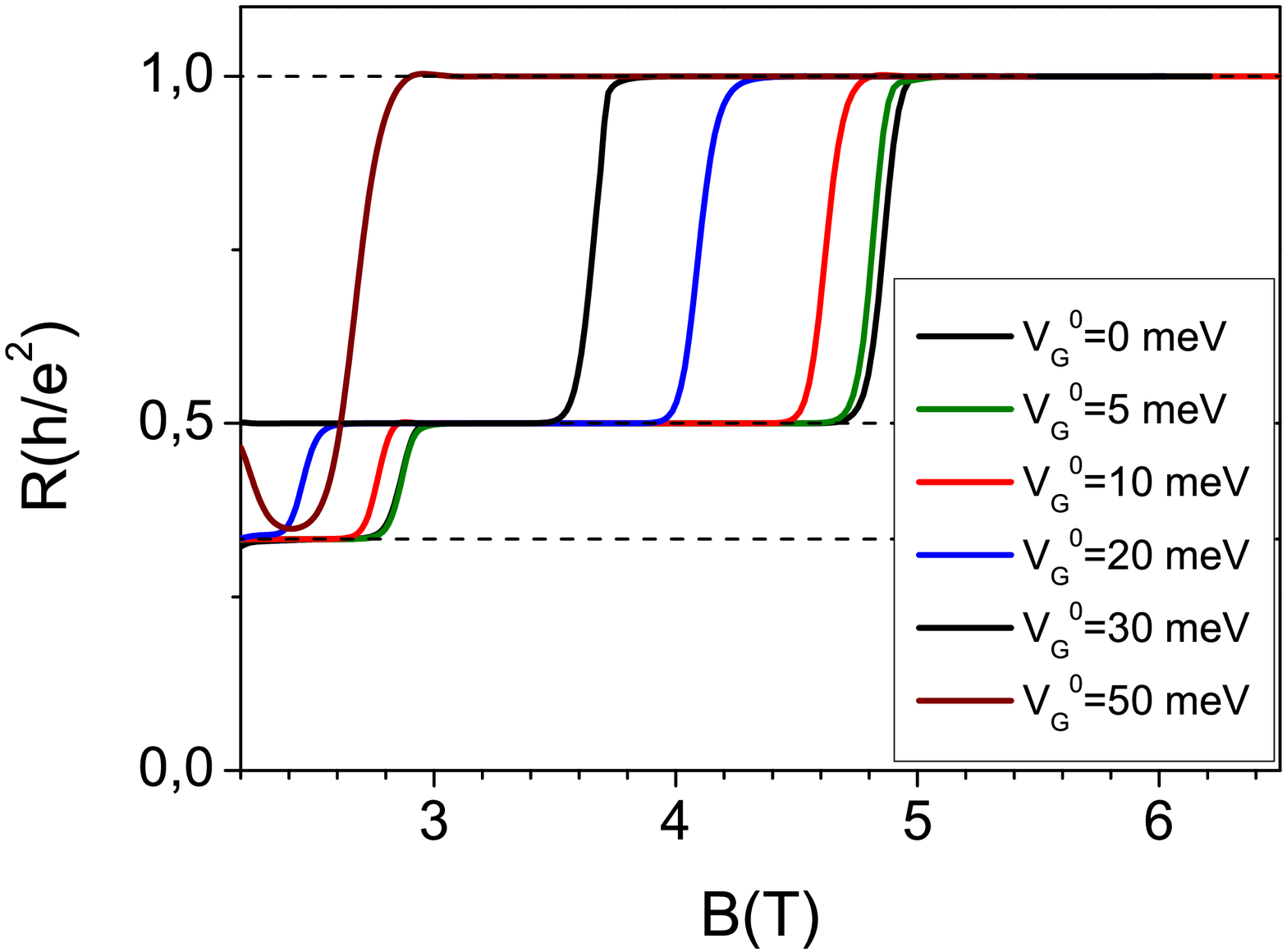}						\\
		\includegraphics[width=0.5\linewidth]{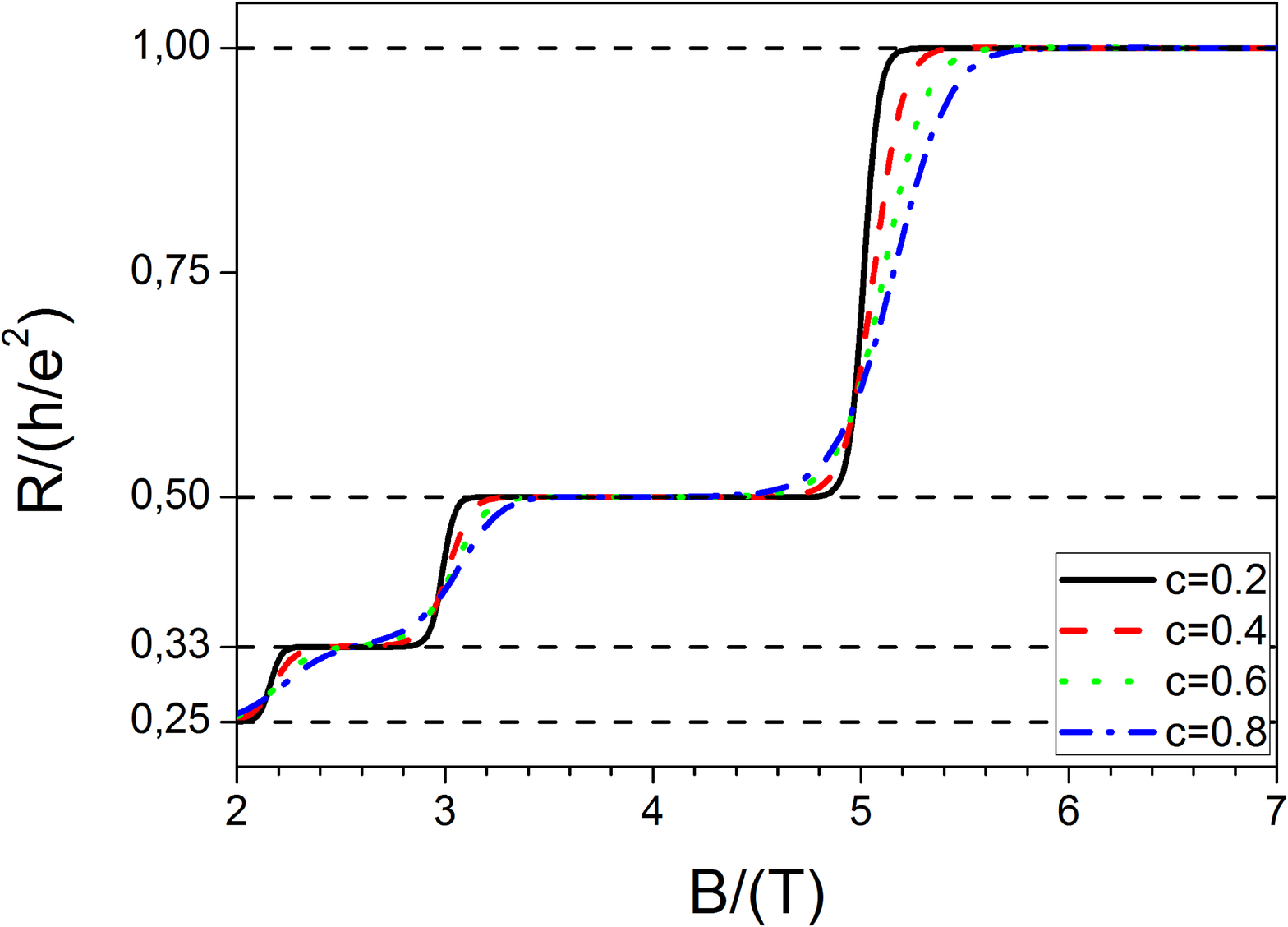}
	\caption{\label{F_Rxy} (color online) Variation of the transversal resistances as a function of 
		magnetic field: The upper plot shows $R_{xy}$ for various gating potentials applied to the 
		inner gated bar. The lower curves show transversal resistance curves for various curvatures c 
		and no gating.}	
\end{figure}

Figure~\ref{F_Rxy} shows $R_{xy}$ for various $V_g^0$ applied to the inner Hall bar. The dominant feature of gating lies in a shift of the plateau transitions to lower magnetic fields due to a reduced electron concentration.
We note that the Fermi energy is fixed by a reservoir and therefore only a function of the magnetic field but not the local gating. The impression of a more or less rigid shift is a consequence of $E_f(B) - V_g(B)$ having small variations on the scale of the magnetic field interval of $R_{xy}$ transitions. We mention that if the gating is large enough, which is the case for $V_g^0=50meV$, the center can jump down to the next lower linear increasing part of the Fermi energy, corresponding to the LL band at lower energy. 

In order to obtain the potential curvatures for all gate potentials, we would have to fit the curvature of each transition and for each $V_g^0$ separately, which is very demanding. Therefore we determine $c_0$ using the ungated $R_{xy}$ curve and calculate from it the respective value $c_v$ for $V_g^0$. 
We focus on the three transitions with lowest filfactor $\nu$, starting with center at highest magnetic field and ordered due to decreasing magnetic field (corresponding to the states $(\nu,s) \in \{(2,+),(2,-),(1,+)\}$ crossing the Fermi energy, where $s$ is the spin orientation). We summarize the manually fitted $c_0$ values as a function of $V_g^0$ in table \ref{T_c0_Vg_fit}.

\begin{table}
	\begin{tabular}{c|c|c|c}
		transition 	& $V_g^0 = -450mV$	& $V_g^0 = -300mV$	& $V_g^0 = -150mV$ \\
		\hline
		$2\to 1$ 		&	0.5							& 0.85						& 1.0		\\
		$3\to 2$ 		&	0.5							& 0.45						& 0.6		\\
		$4\to 3$ 		&	0.3							& 0.4 						& 0.2		\\
	\end{tabular}
	\caption{\label{T_c0_Vg_fit} Fitted curvatures $c_0$ in units $meV/T$ for various transitions and bare 
		gating potentials. We estimate the fitting error as $\pm 0.2 meV*T$ from the line thickness of the 
		Hall resistance in the experimental figure.}
\end{table}

It is interesting to also present the $c_0$ values at the leg without gating, summarized in table \ref{T_c_noGating}.
For each transition the values should not depend on $V_g^0$, if there is no charge transfer from the gated to the ungated leg. It seems that the curvature decreases (the potential landscape is getting flatter) with increased gating on the opposite leg. On the other hand, we expect the charge transfer to be small in such a macroscopic bar. Within the (large) uncertainties of the experimental curves we cannot predict even a qualitative trend. Therefore we average along each row to obtain $0.17$, $0.4$ and $0.67$ in units of $meV/T$ for $4\to 3$, $3\to 2$, and $2\to 1$, respectively.

\begin{table}
	\begin{tabular}{c|c|c|c}
		transition 	& $V_g^0 = -450mV$	& $V_g^0 = -300mV$	& $V_g^0 = -150mV$ \\
		\hline
		$2\to 1$ 		&	0.6						& 0.6						& 0.8 		\\
		$3\to 2$ 		&	0.45					& 0.35					& 0.4 		\\
		$4\to 3$ 		&	0.1						& 0.2						& 0.2 		\\
	\end{tabular}
	\caption{\label{T_c_noGating} Curvatures $c_0$ in units $meV/T$ for various transitions and bare 
		gating potentials for the part of the anti-Hall bar with no gating applied}
\end{table}

We stress that the shift of the transition center with gating also changes the slope of $R_{xy}$ in the center of transitions ($\epsilon=0$) due to the appearence of $B/c$ (note $c=a/2$, to be consistent with the definition of $a$ in \cite{Uiberacker2009}) in $P$ (see Eq.~\ref{E_def_P}). In order to compare transitions for various gatings, we demand that the interval in magnetic field $\delta B_v$, corresponding to $|\epsilon(B) B/c_v|=1$, is the same as the one at zero gating, given by $|\epsilon(B) B/c_0|=1$, using an appropriate definition of $c_v(c_0)$. 
In order to get explicit expressions we expand linearly around the magnetic field at transition centers, $\epsilon(B)\approx \epsilon_0 + \epsilon_1 B_{tr}$. In this way we get from $|\epsilon(B) B/c|=1$ the interval
\begin{equation}
	\delta B = -\frac{B_{tr}}{2} \pm \sqrt{ \frac{B_{tr}^2}{4} + \frac{c}{\epsilon_1} } \quad . 
\end{equation}
Demanding $\delta B_v = \delta B_0$ we map the two transitions onto one curve and arrive at the curvature of the gated sample,
\begin{equation}
	c_v = \epsilon_1^v \left[ -\frac{(B_{tr}^v)^2}{4} + \left( \frac{B_{tr}^v-B_{tr}^0}{2} + \sqrt{ \frac{(B_{tr}^0)^2}{4} + \frac{c_0}{\epsilon_1^0} } \right)^2 \right] \quad . 
	\label{E_cGating}
\end{equation}

Using the values of $c_0$ in table \ref{T_c0_Vg_fit}, we calculate $c_v$ for each transition in terms of Eq.~\ref{E_cGating}, which we summarize in table \ref{T_ceff}.
\begin{table}
	\begin{tabular}{c|c|c|c|c}
		transition 	& $V_g^0 = -450mV$	& $V_g^0 = -300mV$	& $V_g^0 = -150mV$ & $V_g^0 = 0mV$\\
		\hline
		$2\to 1$ 		&	0.25				& 0.51			& 0.93			&	0.67		\\
		$3\to 2$ 		&	0.17				& 0.17			& 0.47			&	0.40		\\
		$4\to 3$ 		&	0.11				& 0.24			& 0.20			& 0.17		\\
	\end{tabular}
	\caption{\label{T_ceff} Effective curvatures $c_v$ in units $meV/T$ for various transitions and bare 
		gating potentials used in the experiment}
\end{table}

\subsection{Potential landscape}

To determine both the amplitude and lengthscale of the potential fluctuations we need another quantity besides the curvatures. In this respect a model of charge fluctuations by statistically independent electrons is very useful \cite{Wulf1988,Gudmundsson1987}. As is well known, if a collection of statistical objects with identical properties (described by the same random variable) are independent then the relative fluctuations of $X:=\sum_j^N X_j$ are given by $\Delta X/<X>\propto N^{-1/2}$, where $<X>$ denotes the average of $X$ and $\Delta X:=\sqrt{<(X-<X>)^2>}=[\sum_{j,k}^N <X_jX_k> - <X>^2 ]^{1/2}/N$ together with $<X_jX_k>=\delta_{jk}$. 

We partition the sample in $N$ cells of equal size and interpret the electron distribution (or number of electrons) in each cell as a random variable. Assuming charge neutrality on the average the absolute fluctuations of the charge density $n$ are then $\delta n = n/N^{1/2}$. Employing the Poisson equation we then get the estimate $V[meV] = \left| \frac{2\pi e 10^3}{\kappa K}\delta n \right|$ for the magnitude of the (long range) potential energy fluctuations \cite{Wulf1988}. $\kappa = 1.3797*10^{-9}C/Vm$ denotes the dielectric constant of the sample (GaAs) and $K$ is the modulus of the smallest wavevector supported by geometry.

The detailed description of the experimental setup \cite{Mani1997apl} lets us estimate the length of each leg as $l = 2mm$ and the transversal width as $w = 0.2mm$. Assuming variations of charge only normal to equipotential lines the largest wavelength should occur in the middle of the plateau transition where the Hall angle is close to $\pi/4$. We arrive at a wavelength of $2\pi/K = 2*10^{-4}\sqrt{2}$ meters. Moreover we get $\delta n = n/\sqrt{N} = \sqrt{nN_c/lw}$. Noting that $n$ hardly varies between transitions we use averages for each $V_g^0$. The  fitted densities (in $10^{-11}cm^{-2}$) turn out to be $n = 1.81$, $n = 1.62$, $n = 1.25$, and $n = 0.88$ for bare gating (in $meV$) $V_g^0 = 0$, $V_g^0 = 10$, $V_g^0 = 30$, and $V_g^0 = 50$, respectively. $N_c$ denotes the number of (correlated) electrons in each fictitious cell of the statistical model. The amplitudes of potential fluctuations become $V/\sqrt{N_c} = 2.203$, $1.972$, $1.521$, and $1.071$, respectively in units 
$meV$, for increasing gating. 

This gives the interesting possibility to get a value for the number of correlated electrons, as explained in the following:
A plausible upper bound for $N_c$ is given by the energy difference between LLs, because the screening tries to suppress higher potential amplitudes due to Wulf et.~al.~\cite{Wulf1988}. Namely, if the potential energy exceeds the LL spacing a new LL band is occupied, which leads to strong screening that suppresses the potential until the new LL band is emptied and screening is weak again. Clearly this argument does not hold for potentials that are so large as to change the level structure significantly or lead to the breakdown regime.
We arrive in this way at the upper bounds $N_c^u$ that are presented in table \ref{T_NcBounds}. The variation with $V_g^0$ turns out to be small.

\begin{table}
	\begin{tabular}{c|c|c|c|c}
		transition 	& $V_g^0 = -450mV$	& $V_g^0 = -300mV$	& $V_g^0 = -150mV$ & $V_g^0 = 0mV$ \\
		\hline
		$2\to 1$ 		&	5						& 5					& 4				& 4	\\
		$3\to 2$ 		&	3						& 2					& 2				& 2	\\
		$4\to 3$ 		&	2						& 2 				& 2				& 2	\\
	\end{tabular}
	\caption{\label{T_NcBounds} Number of correlated electrons in statistically independent cells. We rounded to the next higher/lower integer.}
\end{table}

Using the definition of the curvature we get the potential correlation length in $nm$ as
\begin{equation}
	L = \sqrt{\frac{hV}{e c_v}} = v\sqrt{ \frac{ \sqrt{N_c} }{c_v} }  \quad ,
\end{equation}
where $V = ev^2 \sqrt{N_c}/h$ with numerical values $v = 95.46$, $85.44$, $65.93$, and $46.41$ for increasing gating. The so obtained potential correlation lengths are summarized in table \ref{T_VCorrL}. We judge the error as coming predominantly from the curvature, because the concentration $n$ can be fitted with high accuracy. The resulting error varies strongly due its proportionality to $c^{-3/2}$. Qualitatively the error is less for less gating and smaller fillfactor (that is, for larger value of $L$ in \ref{T_VCorrL}).

\begin{table}
	\begin{tabular}{c|cccc}
		transition 	& $V_g^0 = -450mV$	& $V_g^0 = -300mV$	& $V_g^0 = -150mV$ & $V_g^0 = 0mV$\\
		\hline
		$2\to 1$ 		&	$138.8 \pm 55.5$	& $138.0 \pm 27.1$	& $125.3 \pm 13.5$	& $164.9 \pm 24.6$				\\
		$3\to 2$ 		&	$148.2 \pm 87.1$	& ($190.3 \pm 111.9$)	& $148.2 \pm 31.5$	& $179.5	\pm 44.9$			\\
		$4\to 3$ 		&	($166.5 \pm 151.3$)		& $160.1 \pm 66.7$	& $227.1 \pm 113.6$	& ($275.3	\pm 162.0$)	\\
	\end{tabular}	
	\caption{\label{T_VCorrL} Correlation length and their errors of potential energy fluctuations in $nm$, using the values of $N_c$ from table \ref{T_NcBounds}. Brackets indicate values with large errors.}
\end{table}

Table \ref{T_VCorrL}, which shows microscopic information that is hard to be addressed by experiment, is our main result. These values should be viewed as the largest correlation lengths appearing in the potential landscape, due to the Hartree-like estimation of the potential amplitude used. The general trend is decreasing screening (smaller $L$) with increasing $V_g^0$. This makes sense, despite the high error for low $L$, and is due to decreased carrier concentration. Similar the screening increases with the number of electrons present at transitions, that is with decreasing magnetic field, due to the same reason. However, there are exceptions to the rule which seem to come from the non-trivial dependence of the Fermi energy and screened gating potential energy with the magnetic field.

We stress that via $K$ a larger area of the 2dEG sample leads in principle to higher potential fluctuations, which gives different $N_c$ due to the saturation at the LL energy difference.
Furthermore, the presented (large scale) correlation lengths have a realistic order of magnitude when compared to experiments \cite{McCormick1999,Weitz2000,Ahlswede2001} as well as to our own Hartree-Fock (HF) calculations, using a refined version, with Broyden mixing for fast convergence, of a code originally written by Römer et.~al. \cite{Romer2008}. We note here that in reality due to imperfect screening a reminescence of the distribution of charge from doping centers near the 2dEG affects the assumed statistical independence slightly. However, as shown by Gudmundsson et.~al.~\cite{Gudmundsson1987} the agreement with experiments is good. 

\subsection{Chemical Potential distribution}

\begin{figure}[htbp] %
  \centering
%	B=3.66T   
		\includegraphics[width=0.4\linewidth]{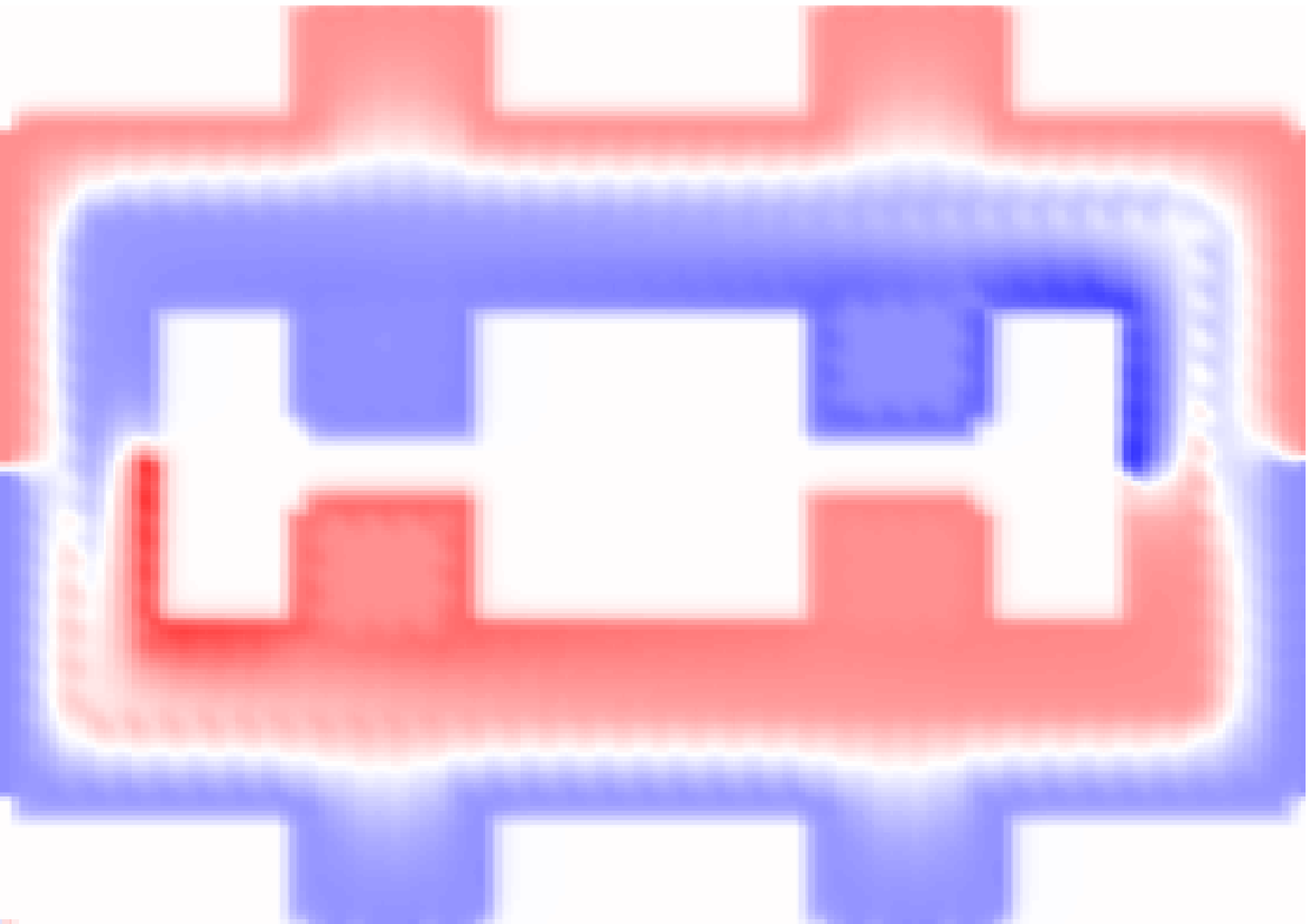} \qquad
% 	B=4.20T
		\includegraphics[width=0.4\linewidth]{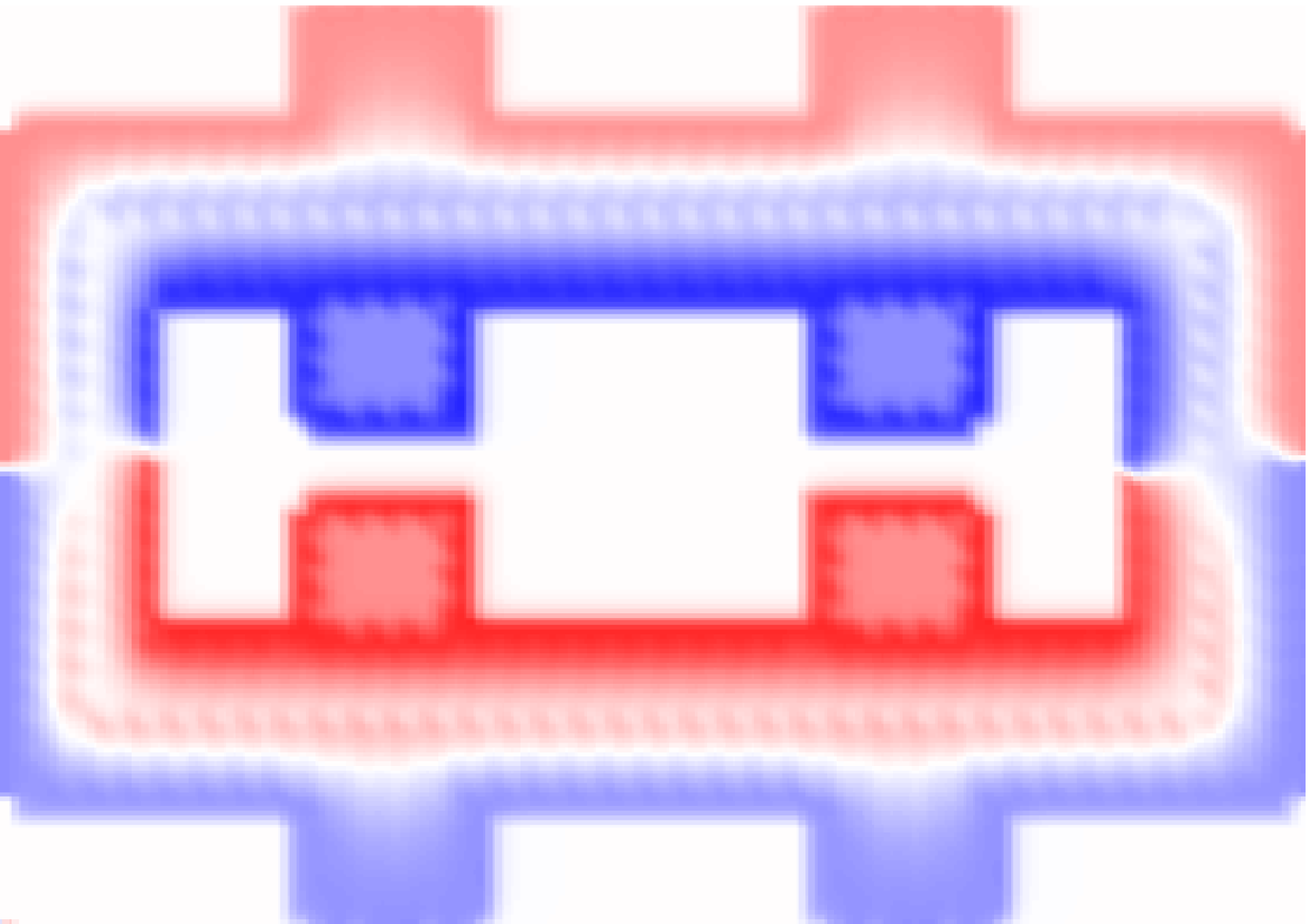} \\ 
% 	B=4.86T
		\includegraphics[width=0.4\linewidth]{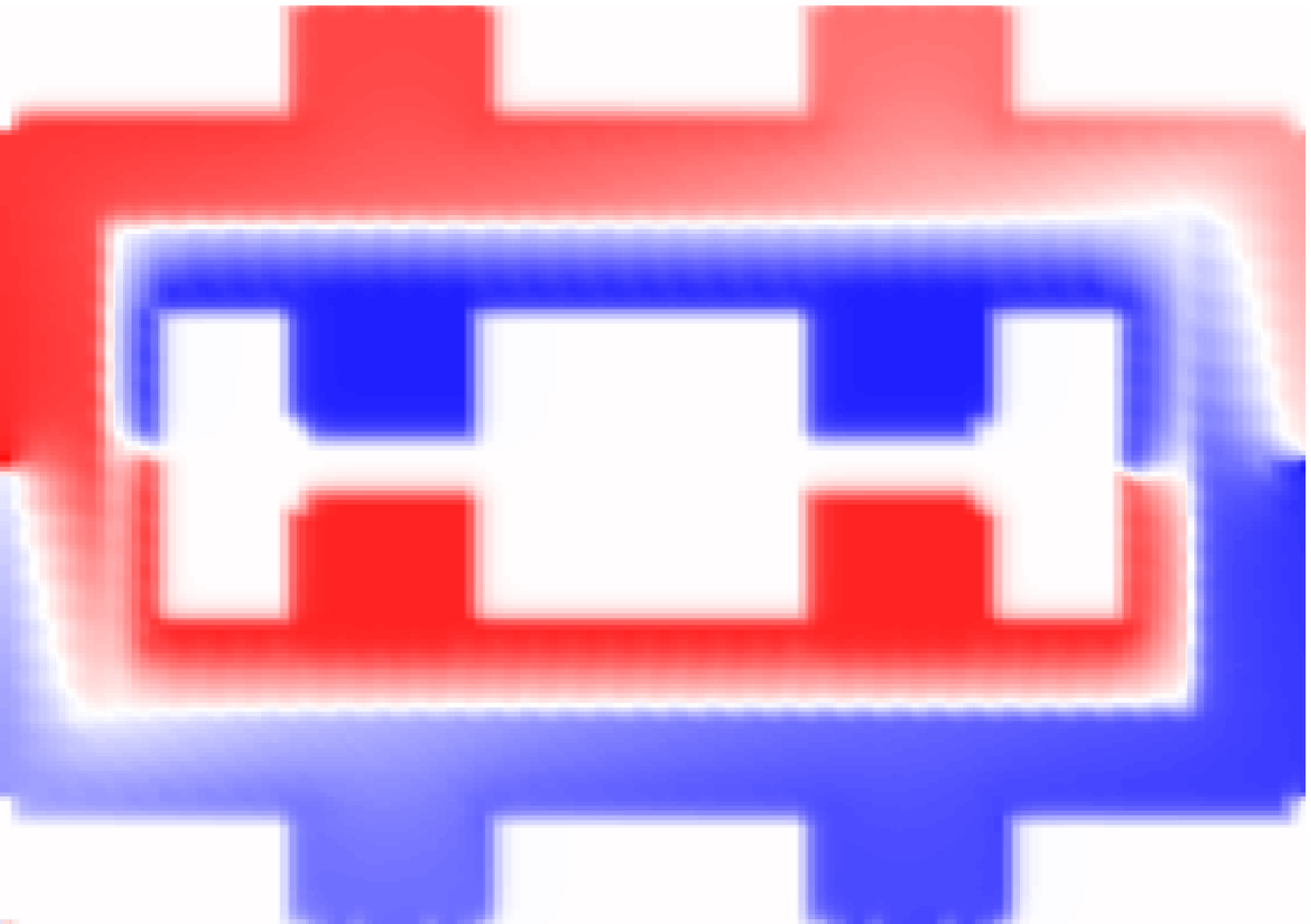}	\qquad
% 	B=5.50T
		\includegraphics[width=0.4\linewidth]{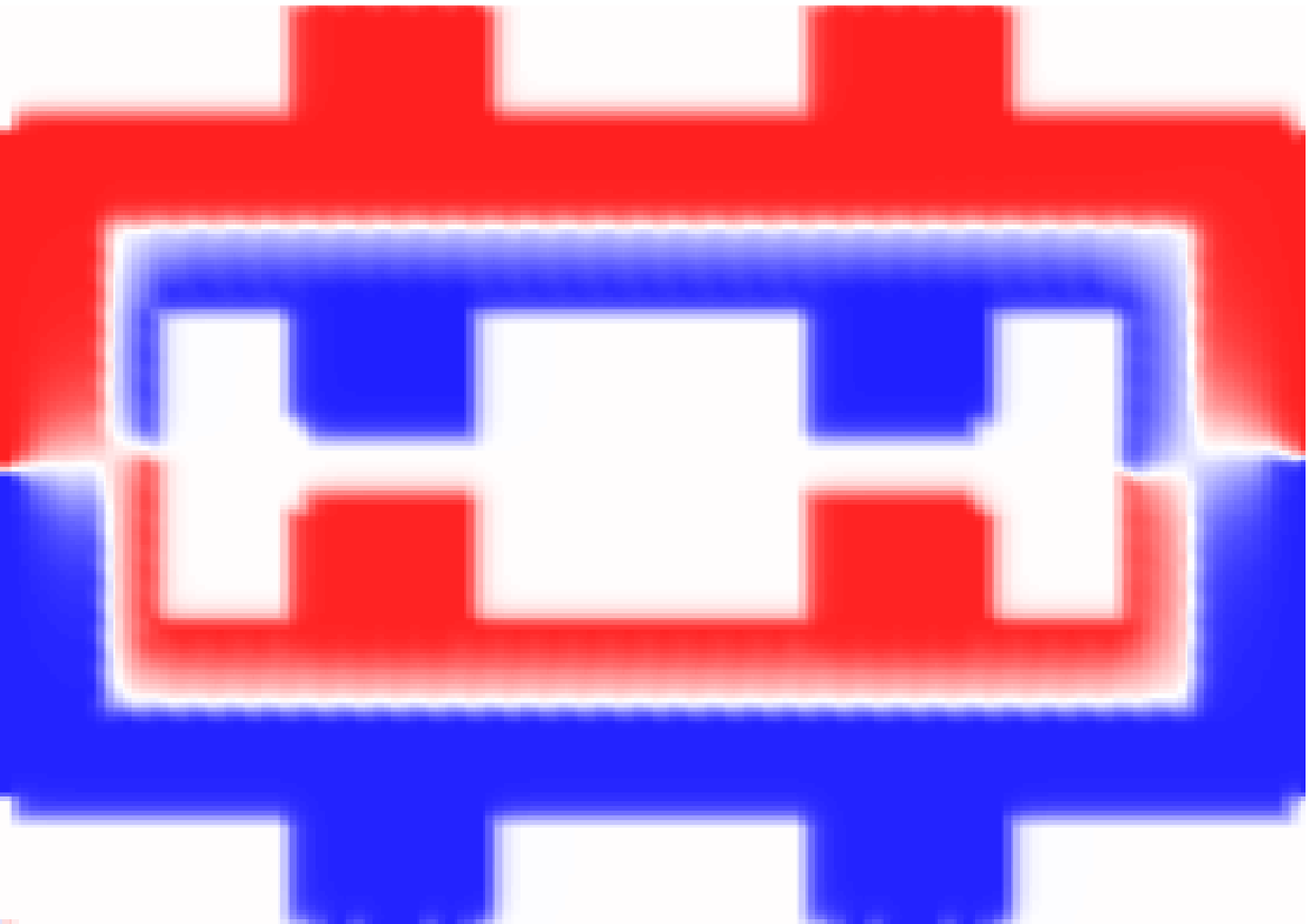}	\\ 
 		\includegraphics[width=0.15\linewidth]{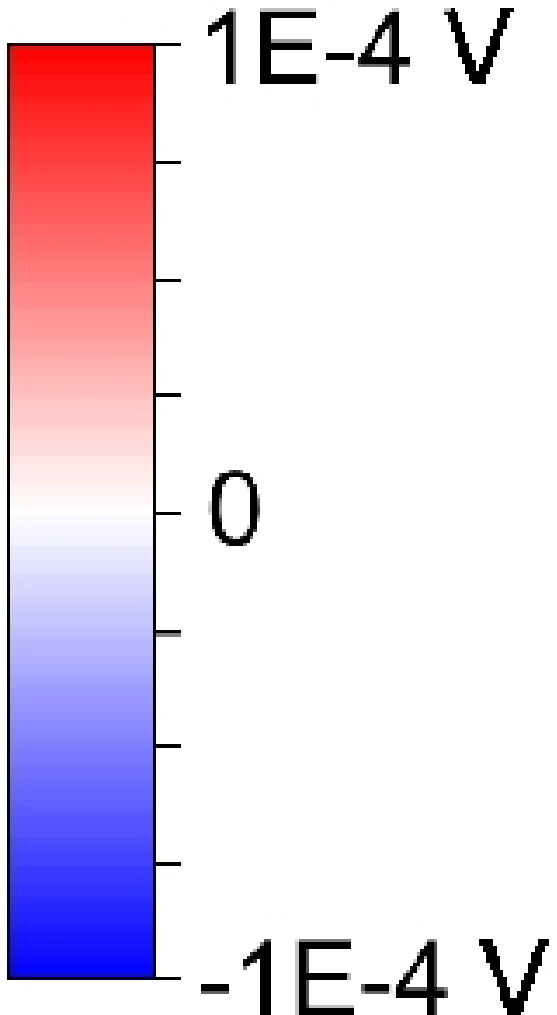}
 \caption{\label{F_ChemPot}(color online) Distribution of the electrochemical potential: we show 
 	results for a bare gating potential of $30meV$ and magnetic field values (from top, left to bottom,
 	right) of $3.66T$, $4.20T$, $4.86T$ and $5.50T$.}
\end{figure}

Finally, in Figure \ref{F_ChemPot} we present details of the electrochemical potential distribution for the case corresponding to a bare gating of \(V^0_G = 30 meV\) ($-300mV$ in the experiment). We selected 4 magnetic field values to show the generic behaviour. We first note that with increasing $B$ the maximum voltage difference in the system increases, which is a consequence of constant injected current. At $3.66T$ the inner (gated) bar is in the transition region, which can clearly be seen from the longitudinal gradient in blue/red color in the upper/lower leg. On the other hand the red color on the other bar is homogeneous, that is, in the plateau and dissipation only occurs at the current contacts.
At $4.20T$ inner and outer bar are in the plateau, therefore only a transversal gradient develops and the line of zero potential (white) is parallel to the longitudinal direction along the leg. $4.86T$ shows qualitatively the same picture as $3.66T$ with the role of inner and other bar reversed. Now the outer bar shows a longitudinal gradient. Finally at $5.50T$ both bars are in the plateau again. 

\section{Summary and Conclusion}
\label{s_conclusion} 

In summary we present simulations of experiments done on an anti-Hall bar within a Hall bar geometry by Mani \cite{Mani1997apl}, which show a shift of the Hall resistance to lower magnetic field by applying partial gating. We calculate the chemical potential distribution for the integer quantum Hall regime under a constant current condition, as used in the experiment. The so obtained transversal resistances as a function of magnetic field compare well with the experimental curves. 

Fitting by hand the position of plateau transitions of the transversal resistance for applied gating, we are able to obtain the electron density, the enhancement factor for the Zeeman interaction, and the screened gating potential in the 2d electron gas. Finally we determine the curvature of potential saddles by fitting the width of plateau transions using the ungated Hall resistance curve. Employing a model of partitioning the sample into statistically independent cells with correlated electrons in each cell, we arrive at the amplitude of potential fluctuations. Defining a transformation we obtain plateau transition intervals for the gated system from the fitted values of the ungated system. In this way we are able to provide the magnitude \textit{and} lengthscale of potential fluctuations from charge inhomogenities as a function of fillfactor of the transition and bare gating potential.

\begin{acknowledgments}                                                                
	This work was sponsored by the Austrian Science Fund under the Research Program 
	No.~P 19353-N16. 
\end{acknowledgments}  

%%%%%%%%%%%%%%%%%%%%%%%%%%%%%%%%%%%%%%%%%%%%%%%%%%%%%%%%%%%%%%%
\bibliography{../../../../Citations/QuantumHallEffect/CitationsQHE}
%%%%%%%%%%%%%%%%%%%%%%%%%%%%%%%%%%%%%%%%%%%%%%%%%%%%%%%%%%%%%%%

\newpage

%\setlength{\parindent}{0cm}
%\setlength{\parskip}{0.5cm}

%%%%%%%%%%%%%%%%%%%%%%%%%%%%%%%%%%%%%%%%%%%%%%%%%%%%%%%%%%%%%%%%
%{\bf Figure captions}
%
%Figure 1. 
%
%Figure 2.  
%
%Figure 3. 
%
%Figure 4.  
%
%Figure 5. 
% 
%\newpage  

%%%%%%%%% 1 %%%%%%%%%%%%%%%%%%%%%%%%%%%%%%%%%%%%%%%%%%%%%%%%%%%%%%

%%%%%%%%%% 2 %%%%%%%%%%%%%%%%%%%%%%%%%%%%%%%%%%%%%%%%%%%%%%70%%%%%%%

%%%%%%%%%% 3 %%%%%%%%%%%%%%%%%%%%%%%%%%%%%%%%%%%%%%%%%%%%%%70%%%%%%%

%%%%%%%%%% 4 %%%%%%%%%%%%%%%%%%%%%%%%%%%%%%%%%%%%%%%%%%%%%%70%%%%%%%
\begin{figure}[htbp] %
  \centering
		\includegraphics[width=0.5\linewidth]{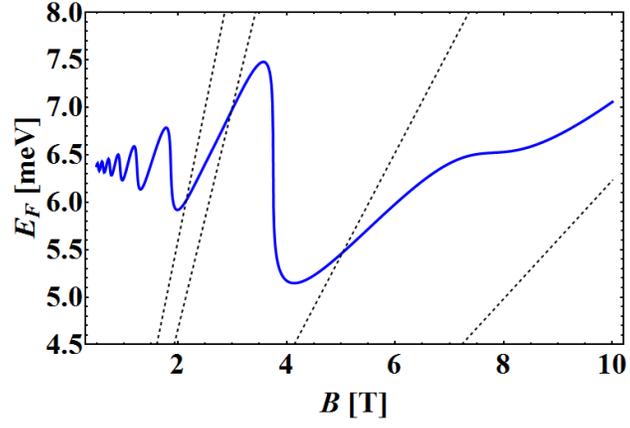}
 \caption{\label{F_EFermi}(color online) Fermi energy and 4 Landau levels of lowest energy 
 	(2 spin-resolved pairs) as functions of the magnetic field.}
\end{figure}
%%%%%%%%%% 5 %%%%%%%%%%%%%%%%%%%%%%%%%%%%%%%%%%%%%%%%%%%%%%70%%%%%%%

%%%%%%%%%%%%%%%%%%%%%%%%%%%%%%%%%%%%%%%%%%%%%%%%%%%%%%%%70%%%%%%%

\end{document}